\documentclass[iop,apj]{emulateapj}
\usepackage{natbib}
\usepackage{xcolor}
\usepackage{hyperref}
\hypersetup{
   colorlinks,
   linkcolor={blue!88!black!80},
   citecolor={blue!88!black!80},
   urlcolor={blue!88!black!80}
}

\begin{document}

\title{More than softer-when-brighter: the X-ray powerlaw spectral variability in NGC 4051}

\author{Yun-Jing Wu$^{1,3}$, Jun-Xian Wang$^{\star1,2}$, Zhen-Yi Cai$^{1,2}$, Jia-Lai Kang$^{1,2}$, Teng Liu$^{4}$, Zheng Cai$^{3}$}
\affil{
$^1$CAS Key Laboratory for Research in Galaxies and Cosmology, Department of Astronomy, University of Science and Technology of China,
Hefei, Anhui 230026, China; yj-wu19@mails.tsinghua.edu.cn, *jxw@ustc.edu.cn\\
$^2$School of Astronomy and Space Science, University of Science and Technology of China, Hefei 230026, China\\
$^3$Department of Astronomy, Tsinghua University, Beijing 100084, China\\
$^4$Max-Planck-Institut f$\rm\ddot{u}$r extraterrestrische Physik, Giessenbachstrasse 1, D-85748 Garching bei M$\rm\ddot{u}$nchen, Germany
}

\begin{abstract}
The powerlaw X-ray spectra of active galactic nuclei at moderate to high accretion rates normally appear softer when they brighten, for which the underlying mechanisms are yet unclear. Utilizing XMM-Newton observations and  excluding photons $<$ 2 keV to avoid contamination from the soft excess, in this work we scrutinize the powerlaw spectral variability of NCG 4051 from two new aspects. 
We first find that a best-fit ``softer-when-brighter'' relation is statistically insufficient to explain the observed spectral variabilities, and intervals deviated from the empirical relation are clearly visible in the light curve of 2 -- 4 keV/4 -- 10 keV count rate ratio.  
The deviations are seen not only between but also within individual XMM-Newton exposures, consistent with random variations of the corona geometry or inner structure (with timescales as short as $\sim$ 1 ks), in addition to those behind the smooth ``softer-when-brighter'' trend.
We further find the ``softer-when-brighter" trend gradually weakens with the decreasing timescale (from $\sim$ 100 ks down to 0.5 ks). These findings
indicate that the powerlaw spectral slope is not solely determined by its brightness.
We propose a two-tier geometry, including flares/nano-flares on top of the inner disc and an embedding extended corona (heated by the flares, in analogy to solar corona) to explain the observations together with other observational clues in literature. Rapid spectral variabilities could be due to individual flares/nano-flares, while slow ones are driven by the variations in the global activity of inner disc region (akin to the variation of solar activity, but not the accretion rate) accompanied with heating/cooling and inflation/contraction of the extended corona.  
\end{abstract}


\section{Introduction} \label{sec:intro}

The hard X-ray radiation of active galactic nuclei is believed to be produced through thermal Comptonization of low energy seed photons from the accretion disc by hot plasma surrounding the
SMBH, presumably the corona \citep{Galeev1979, Haardt1991, Haardt1993}, and the yielded X-ray spectra are commonly described with powerlaw cut-offed at high energy
\citep[e.g.][]{Zdziarski1995,Ricci2011,Tortosa2018}. 

The X-ray emission of AGNs is strongly variable, not only in the flux but also in the spectral shape. Observing the temporal variations of the coronal spectra provides essential opportunities to probe the yet poorly understood corona physics. 
The mostly known X-ray spectral variability is the so-called ``softer-when-brighter" behavior, i.e., the X-ray coronal spectra of AGNs at moderate to high accretion rates usually get softer when they brighten in X-ray \citep[e.g.][]{Markowitz2004,Sobolewska2009,Soldi2014}.
Note, a contrary ``harder-when-brighter" trend is seen in low luminosity AGNs, for which the X-ray production process could be different \citep{Emmanoulopoulos2012,Connolly2016}.

The ``softer-when-brighter" behavior appears qualitatively consistent with the fact that, in the regime of moderate to high accretion rates (the focus of this work), AGNs with higher Eddington ratios tend to have softer X-ray spectra \citep[e.g.][]{Shemmer2006,Risaliti2009,Yang2015}.
It is generally proposed that when an AGN increases its accretion rate, the corona would be cooled down, therefore generating brighter but softer X-ray spectra.  
However, it is hard to attribute rapid X-ray spectral variability in individual AGNs to changes of accretion rates due to the timescale discrepancy (i.e, the timescale of accretion rate variation should be much longer). 
We further note that,  the X-ray emission of AGNs is generally more variable than optical/UV radiation (at least at timescales up to a few weeks, e.g., \citealt{Uttley2006}, and also see \citealt{Alston2013} for NGC 4051, the target of this work), thus a higher fraction of energy is dissipated into the corona when an individual AGN brightens in X-ray;
contrarily, AGNs with higher Eddington ratios tend to have smaller X-ray to bolometric luminosity ratios \citep{Wang2004, Vasudevan2009, Grupe2010, Lusso2012}.
This clearly objects
the hypothesis that the fast ``softer-when-brighter" trend in individual AGNs and the $\Gamma$ -- Eddington ratio correlation within AGN samples could both be attributed to variations of accretion rate.

Alternatively, the rapid X-ray spectral variabilities in individual AGNs could be attributed to variations of the coronal properties. 
For instance, \cite{Zhang2018} found with a small sample of sources that AGNs tend to have larger hard X-ray cutoff energies (thus hotter corona) when they brighten in X-ray. This is
in direct contrast to the corona cooling hypothesis aforementioned, and requires a smaller corona opacity at brighter phases to produce the softer spectra \citep[e.g.][]{Keek2016}.
Furthermore, \cite{Sarma2015} found that, in the X-ray photon index $\Gamma$ versus luminosity $L_X$ plot, Mrk 335 follows a rather different track during one XMM-Newton exposure compared with other exposures. This indicates that there is not a canonical ``softer-when-brighter" trend even for a single source, and structural changes in the corona would be required.
More evidence supporting the scheme of non-static corona includes spot-like flares on top of the disc \citep[e.g.][]{Iwasawa2004}, jet-like expanding flares \citep[e.g.][]{Wilkins2015,Alston2020}, vertical outflowing coronae \citep[e.g.][]{Liu2014}, etc.

A fundamental question can then be raised: how fast can such coronal structural changes happen? Identifying the shortest timescale for such variations could uniquely probe the corona physics. 
Meanwhile, rapid and slow X-ray variabilities may involve different physical processes.  Thus a parallel concern is whether the X-ray variabilities at different timescales follow the same ``softer-when-brighter" trend \citep[also see][]{Lobban2018}.

NGC 4051, one of the X-ray brightest AGNs, is a low mass Seyfert 1 galaxy. 
Its X-ray variabilities have been extensively studied in literature \citep[e.g.][]{Lamer2003,McHardy2004,Miniutti2004,Uttley2004,Pounds2004,Ponti2006}.
In this work we scrutinize its hard X-ray spectral variabilities from two new aspects, using archival XMM-Newton observations. 

\section{XMM-Newton observations and data reduction}
In Table \ref{OBS_Info} we list the 19 archival XMM-Newton observations of NGC 4051, all of which were obtained with the ``Medium" blocking filter. In this work we focus on data obtained with the PN detector,  which was operated in Small Window model during all 19 exposures.
We reprocessed the exposures with XMM-Newton Science Analysis System (SAS 17.0.0). The average pile-up effect during individual exposures is found to be negligible with the task ``epatplot". 
\cite{Alston2013MNRAS} examined the pile-up effects within the individual exposures and found only the highest flux revolution (ID: 0606321601) showed signs of pile-up effects during its highest flux periods. Excluding this exposure or its highest flux periods does not alter the major results presented in this work. 
We extract source light curves from a circular region with a radius of 60\arcsec, and background from nearby source-free regions. 
The light curves were obtained using the ``epiclccorr" task and applying both relative and absolute corrections to correct various effects that affect the detection efficiency,
which enables us to directly utilize the derived light curves and their ratios to investigate the rapid spectral variability of NGC 4051.

\begin{deluxetable}{cccc}[b!]
\tablecaption{Archival XMM-Newton observations of NGC 4051. \label{OBS_Info}}
\tablecolumns{4}
\tablenum{1}
\tablewidth{0pt}
\tablehead{
\colhead{Obs. ID.} &
\colhead{Start time(UTC)} &
\colhead{Exp.}  &
\colhead{0.5-10 keV}\\
\colhead{} & \colhead{(YYYY-mm-dd)} & \colhead{(ks)} & \colhead{counts/s}
}
\startdata
0109141401 & 2001-05-16 & 121.958   & 19.99\\
0157560101 & 2002-11-22 & 51.866  & 3.97\\
0606320101 & 2009-05-03 & 45.717 &  7.86\\
0606320201 & 2009-05-05 & 45.645 & 13.69\\
0606320301 & 2009-05-09 & 45.548 &  15.02\\
0606320401 & 2009-05-11 & 45.447 & 4.08\\
0606321301 & 2009-05-15 & 32.644 &  18.86\\
0606321401 & 2009-05-17 & 42.433 &   10.54\\
0606321501 & 2009-05-19 & 41.813 &   13.31\\
0606321601 & 2009-05-21 & 41.936 &   22.02\\
0606321701 & 2009-05-27 & 44.919 &   5.59\\
0606321801 & 2009-05-29 & 43.726 &   7.05\\
0606321901 & 2009-06-02 & 44.891 &   3.75\\
0606322001 & 2009-06-04 & 39.756 &   6.11\\
0606322101 & 2009-06-08 & 43.545 &   2.18\\
0606322201 & 2009-06-10 & 44.453 &   6.22\\
0606322301 & 2009-06-16 & 42.717 &   8.61\\
0830430201 & 2018-11-07 & 83.200 &   11.42\\
0830430801 & 2018-11-09 & 85.500 &   7.18
\enddata
\end{deluxetable}

\begin{figure*}[ht!]
\includegraphics[width=4in,height=3in]{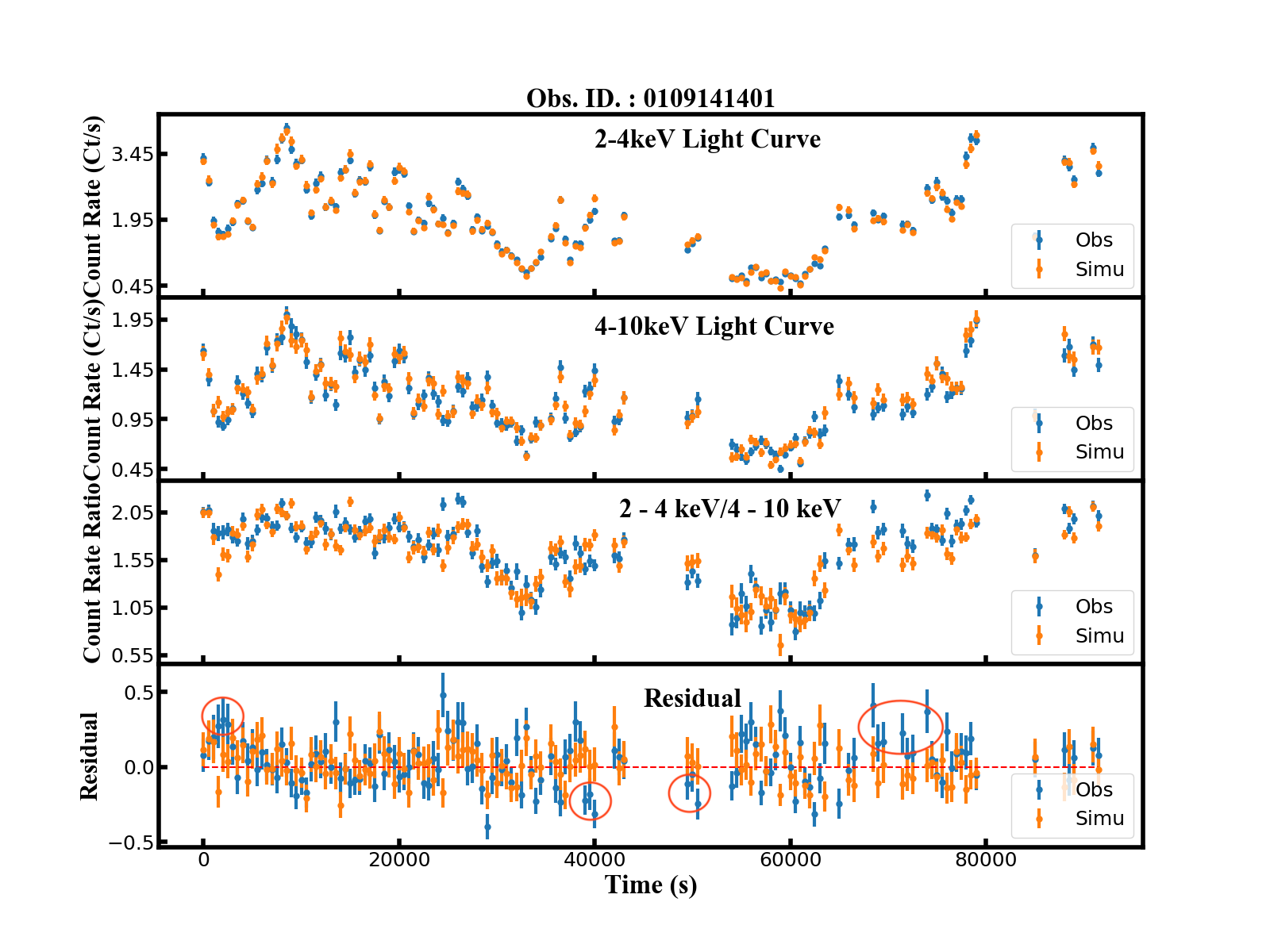}
\includegraphics[width=3in,height=3in]{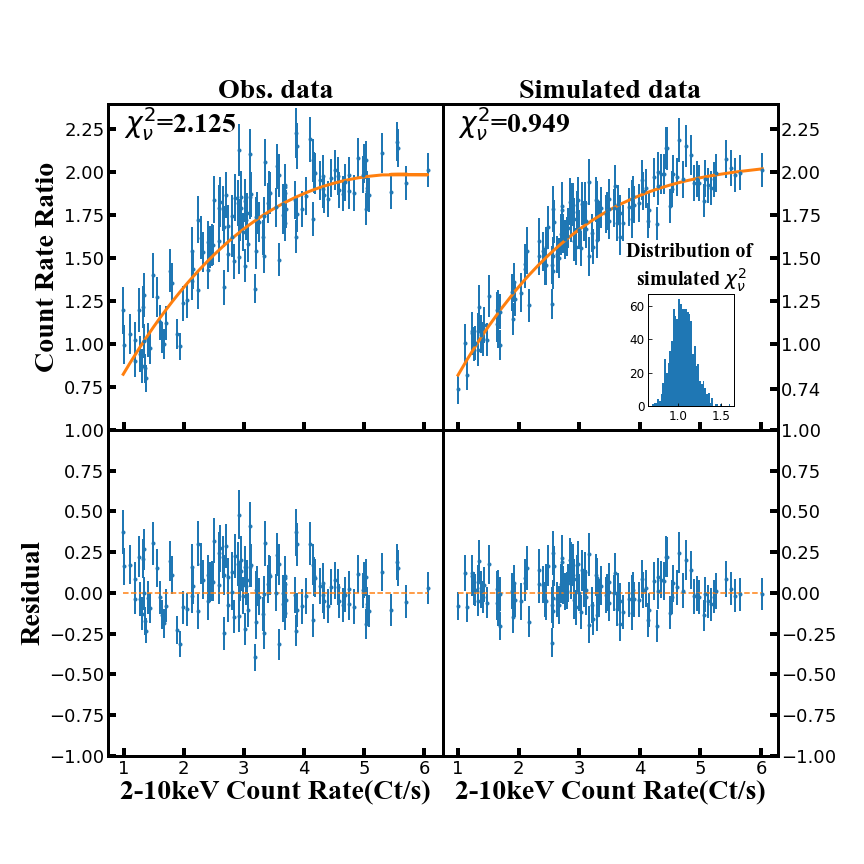}
\caption{Left: XMM-Newton PN light curves of NGC 4051 (ObsID 0109141401, with a bin size of 500s).  The gaps in the light curves are due to high background intervals which had been filtered out during the data processing. Using a looser filter threshold could eliminate some of the gaps but would not alter any of the main results presented in this work.
From top to bottom: 1) 2 -- 4 keV light curves; 2) 4 -- 10 keV light curves; 3) the light curves of 2 -- 4 keV/4 -- 10 keV count rate ratio; 4) the residual count rate ratio light curves after subtracting the empirical relation as shown in the right panel. Simulated light curves which simply follow the empirical ``softer-when-brighter'' relation in the right panel (with statistical errors included) are over-plotted for comparison. Red circles/ellipses mark  example intervals during which significant deviations from the empirical relation are seen. 
Right: The ``softer-when-brighter'' behavior of NGC 4051 seen during this XMM-Newton exposure (left) and for a simulated exposure (right). The solid line plots the best-fit cubic polynomial function, showing an empirical ``softer-when-brighter'' trend (saturated at the bright end), and the yielded $\chi^2$/dof are given. The inserted histogram plots the distribution of expected $\chi^2$/dof if spectral variabilities simply follow the empirical trend.   
\label{f1}}
\end{figure*}

\section{Spectral Variability Analyses}

\subsection{More complicated than an empirical ``softer-when-brighter" relation}
In this work we focus on the spectral range of 2 -- 10 keV to minimize the impact of the soft X-ray excess, and possible variable absorption along the line of sight.
In Fig. \ref{f1} we plot 2--4 keV and 4--10 keV light curves (with bin size  of 500 s) of the first XMM-Newton exposure (ObsID 0109141401), which is also the one with longest effective exposure time ($\sim$ 122 ks).
The 2--4 keV/4--10 keV count rate ratio is also presented, demonstrating rapid spectral variations within this single exposure. 
In the right panel of Fig. \ref{f1} we plot the 2--4 keV/4--10 keV count rate ratio as a function of the  the 2--10 keV count rate. A clear softer-when-brighter trend is seen, and the trend tends to saturate at the bright end (see also \citealt{Seifina2018}).  
We fit the trend with a cubic polynomial function\footnote{We simply adopt this non-parametric approach to derive the smooth ``softer-when-brighter" trend without prior assumption(s) of its shape.
 The results presented in this work however are insensitive to the selection of the function. Adopting other non-parametric approaches such as spline or moving averaging yields similar results.} to empirically describe the ``softer-when-brighter" pattern. While the best-fit empirical relation can well describe the general ``softer-when-brighter'' trend (see the residuals in the right panel of Fig. \ref{f1}), the fit is statistically poor (with $\chi^2$/dof significantly $>$ 1.0), showing the spectral variability is more complicated than the simple empirical relation. 

In Fig. \ref{f1} we mark example intervals when the observed 2--4 keV/4--10 keV count rate ratio deviates from the empirical relation. We see such deviations throughout the whole exposure, i.e., not dominated by a certain interval. The deviations appear stochastic and irrelevant to the total count rate, i.e., not particularly seen during bright or faint states. Furthermore,  the deviations can emerge/disappear on a very short timescale (as short as $\sim$ 2 ks). 
We note that such deviations could also be visible in the flux-flux plots, such as shown in \cite{Taylor2003} for MCG -6-30-15, but the flux-flux plots alone can not reveal the intervals with clear deviations.

\begin{figure*}[ht!]
\includegraphics[width=4.5in,height=3in]{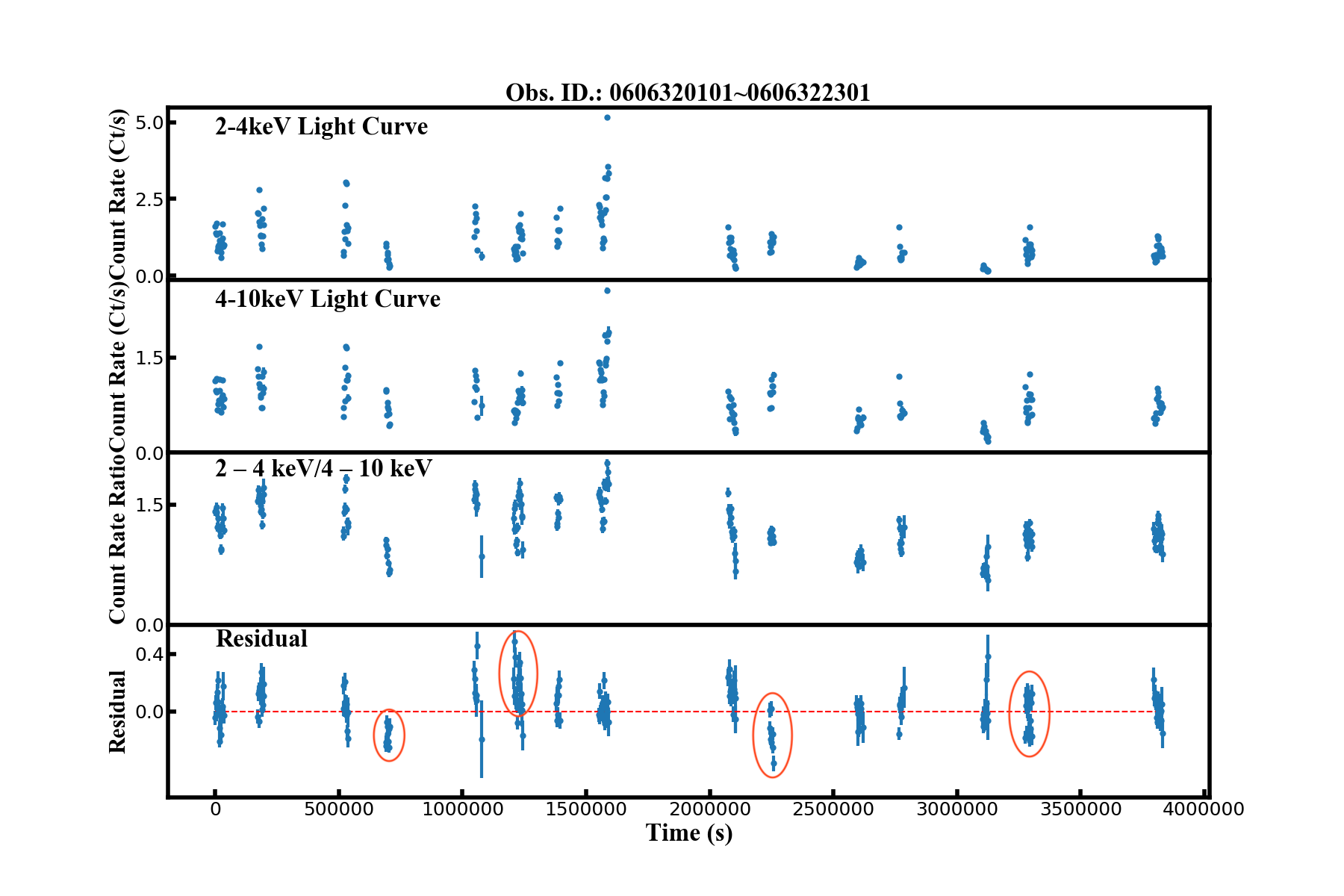}
\includegraphics[width=2.5in,height=3in]{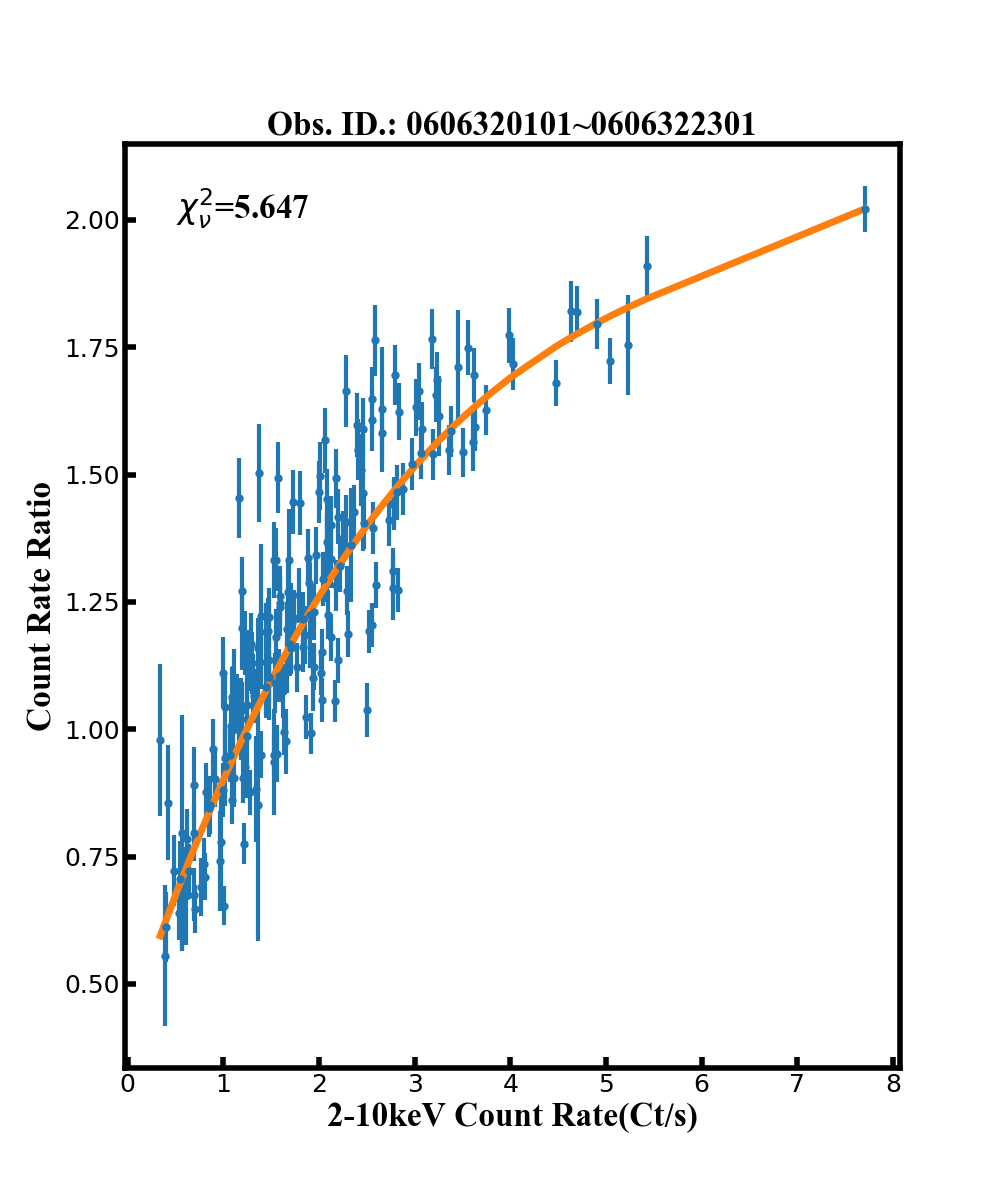}
\caption{
Similar to Fig. \ref{f1}, but for 15 XMM-Newton exposures obtained over 45 days in 2009 (with a bin size of 2,000s). For simplicity, simulated light curves are not shown here. 
\label{f2}}
\end{figure*}

We perform Monte-Carlo simulations to show how the spectral variation would look like if simply following the empirical relation (the spectral slope solely determined by the brightness).
For each time bin, starting from the observed 2--10 keV net count rate, we calculate the expected 2--4 and 4--10 keV net count rates using the  empirical relation.
We then add random Poisson errors to the expected count rates. One set of such artificial light curves is shown in Fig. \ref{f1} (left panel), and the derived
2--4 keV/4--10 keV ratio versus 2--10 keV count rate can be well described with a cubic polynomial function (right panel in Fig. \ref{f1}).
Such simulations have been repeated 1000 times, and the yielded distribution of $\chi^{2}_{\nu}$ is also given in Fig. \ref{f1}. This confirms that 
an empirical relation is insufficient to describe the observed spectral variability.

We repeat above analyses for all 19 individual XMM-Newton exposures, and yield $\chi^{2}_{\nu}$ $>$ 1.2 for 15 exposures, indicating deviations from empirical relations are similarly seen in most exposures. 
We note that among the 19 observations, 15 were obtained quasi-continuously  over 45 days during a monitoring campaign in May -- June, 2009. These observations enable
us to investigate the spectral variability over timescales longer than an individual exposure. The results, similar to that seen in Fig. \ref{f1}, are plotted in Fig. \ref{f2}.
In the plot we can see individual exposures during which the 2--4/4--10 keV ratio significantly deviates from an empirical relation (derived for all 15 exposures).

\subsection{Timescale dependency}

In optical/UV bands, AGNs are known to appear bluer when they brighten in fluxes, which is often described with a similar term ``bluer-when-brighter" (e.g. \citealt{Schmidt2012}). Recent studies have shown that the color variabilities 
in optical/UV are timescale dependent, in the way that the ``bluer-when-brighter" trend is more prominent at shorter timescales,  
and such discoveries provide crucial clues to understand the nature of optical/UV variations \citep{Sun2014, Zhu2018, Cai2016,Cai2018,Cai2019}.
Below we examine whether the X-ray ``softer-when-brighter" trend in NGC 4051 is timescale dependent.
As the 2 -- 4 and 4 -- 10 keV light curves are well coordinated (see Fig.\ref{f1} \& \ref{f2}, also see \citealt{McHardy2004}), the ``softer-when-brighter" trend can be quantified with the ratio 
of proportional variation amplitudes in two bands. For instance, if the source varies 30\% in 2 -- 4 keV but only 20\% in 4 -- 10 keV, the ratio 0.3/0.2
indicates a clear ``softer-when-brighter" trend, the larger the ratio, the stronger the ``softer-when-brighter" trend. A ratio less than unity contrarily corresponds to ``harder-when-brighter".
Following \cite{Zhu2016}, we first derive the structure functions (SFs) of 2--4 keV and 4--10 keV light curves, and utilize the ratio of two structure functions 
to quantify the ``softer-when-brighter" trend as a function of the timescale. 
The structure functions are calculated by using the following formula \citep{diClemente1996}, in which $Log(CR_{i}$) and $Log(CR_{j}$) represent the logarithmic count rates at any two epochs in the light curve,  $\sigma_{i}, \sigma_{j}$ the corresponding logarithmic statistical errors, and $\tau$ the lag of two epochs.

\begin{equation}
SF(\tau) = \sqrt{\frac{\pi}{2}\langle|Log(CR_{i})-Log(CR_{j})|\rangle^{2} - \langle\sigma^{2}_{i}+\sigma^{2}_{j}\rangle}
\label{eq1}
\end{equation}

\begin{figure*}[ht!]
\center
\includegraphics[width=4.5in,height=4.5in]{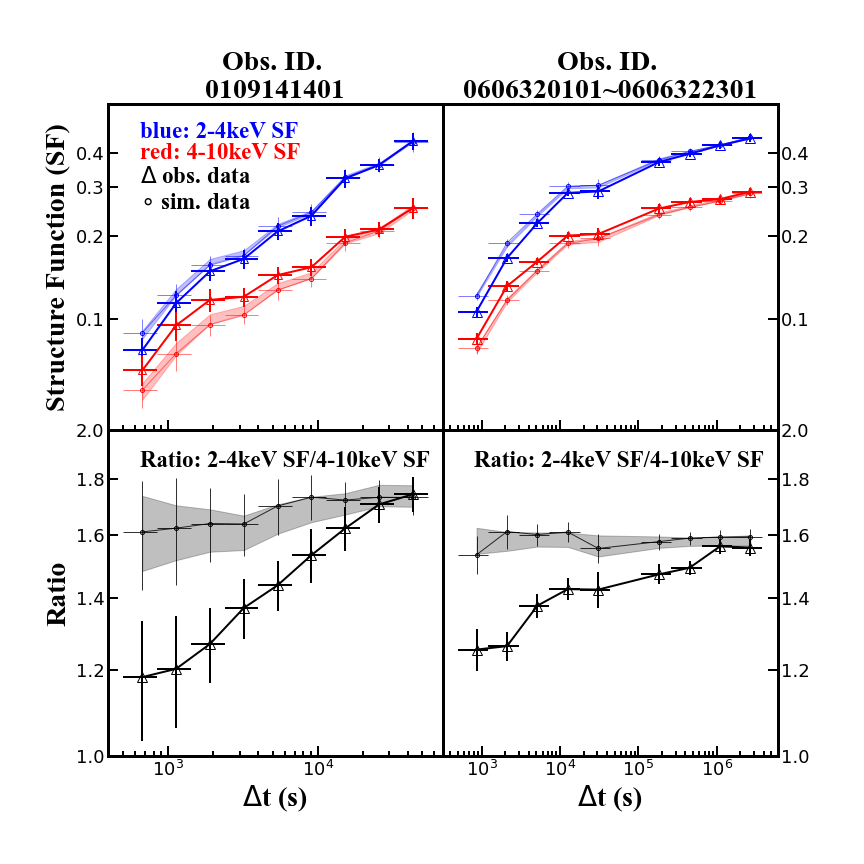}
\caption{2 -- 4 keV and 4 -- 10 keV structure functions (in unit of dex) and their ratios, derived from observed XMM-Newton light curves and simulated ones. Left panels for ObsID 0109141401, and right for ObsID 0606320101 -- 0606322301 which covers a larger range of timescale. The error bars for the structure functions and their ratios are derived through boot strapping the data points in the light curves. 
Measurements from one example simulation are plotted, together with the 1$\sigma$ distribution range with the shaded regions. 
\label{SF_analysis}}
\end{figure*}

Fig. \ref{SF_analysis} plots the derived structure functions, which describe the variation amplitude (in unit of dex) as a function of the timescale.
 The 2 -- 4 keV and 4 -- 10 keV SFs clearly demonstrate that
NGC 4051 is more variable in 2 -- 4 keV than in 4 -- 10 keV within the timescale range of 0.5 ks to 50 ks,  consistent with the ``softer-when-brighter" scenario shown in Fig. \ref{f1} \& \ref{f2}.
However, we find that, the 4--10 keV SFs are flatter than those of 2--4 keV, and the ratios of SFs significantly decrease with the 
decreasing timescale, i.e., the amplitude of the ``softer-when-brighter" trend is timescale dependent. 
The errors in SF($\tau$) and the ratio of SF($\tau$) are obtained through bootstrapping the two light curves \citep{Peterson2001}.
We note while there are known caveats on the use of structure functions (see \citealt[][]{Emmanoulopoulos2010}), e.g., the directly measured structure function may show
spurious breaks at low frequency, in this work what we concern is not the absolute slope of the structure functions, but the difference (ratio) between two bands.
As 2--4 and 4--10 keV light curves  are identically sampled and highly correlated, the biases to SFs due to red noise leakage and aliasing effects should 
have been mostly cancelled out in the ratio of two structure functions (see next paragraph for further simulations). 
For comparison, we also derive the SF($\tau$) and the ratio from the simulated light curves (see \S3.1).
Clearly, the simulated light curves, which solely follow the best-fit smooth ``softer-when-brighter" trends as shown in Fig. \ref{f1} \& \ref{f2},  do not display such timescale dependency. 

\begin{figure}[ht!]
\plotone{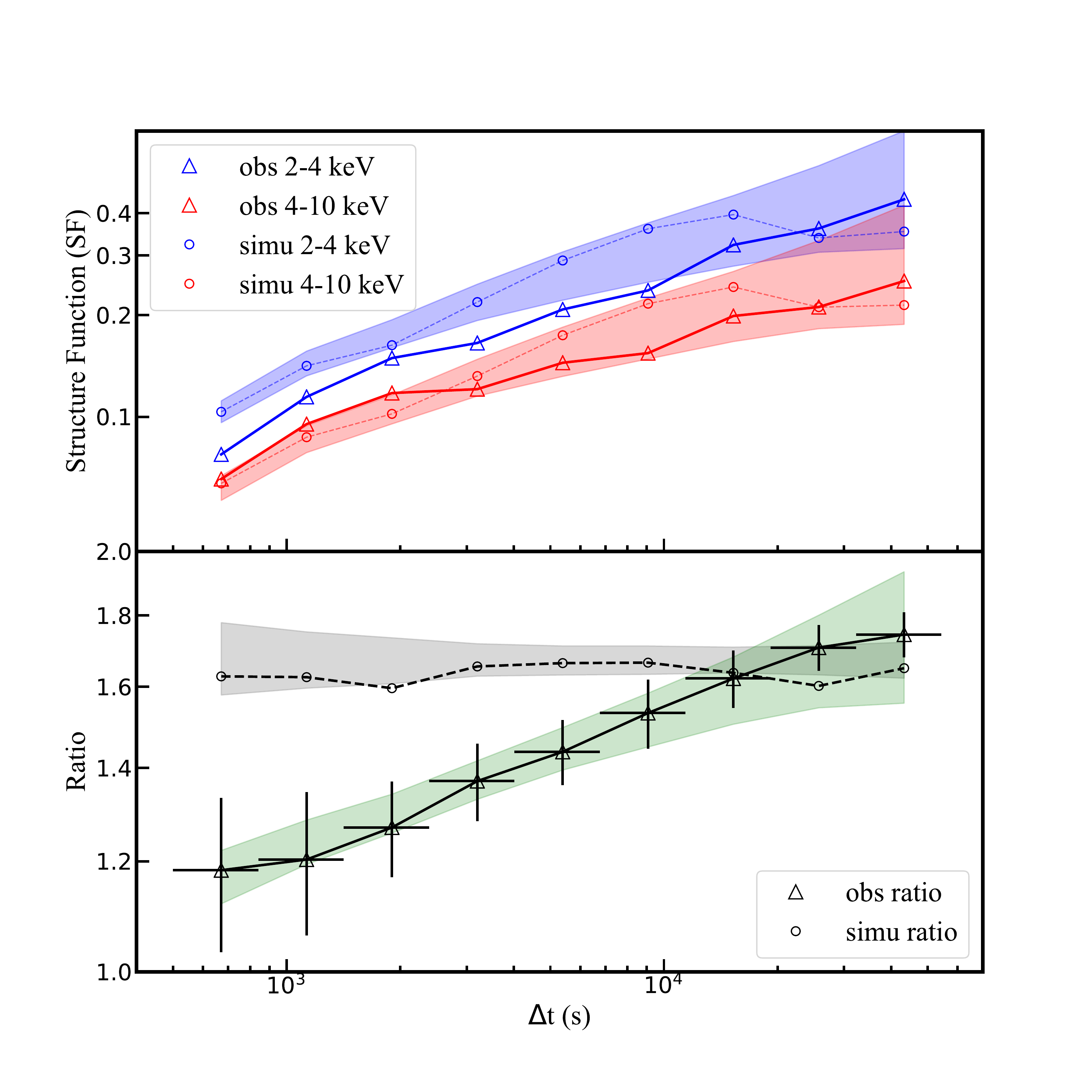}
\caption{
Similar to the left panels of Fig. \ref{SF_analysis}, but here the simulated SFs are derived from PSD-based mock light curves. 
Upper panel:  example simulated SFs for 2--4 keV and 4--10 keV (circles connected with dashed lines) derived from a pair of mock light curves,  
the 16\% - 84\% percentile ranges (blue and red shaded regions) of simulated SFs derived from 1000 pairs of mock light curves,
and for comparison the observed SFs (triangles) from ObsID 0109141401.
The mock light curves are simulated using identical PSD shape ($\alpha$=-1.6) for two bands.
Lower panel: the corresponding SF ratios. The green shaded region plots 16\% - 84\% percentile range of the expected SF ratio from a flatter PSD in 4--10 keV band  ($\alpha_{2-4keV}$=-1.6, $\alpha_{4-10keV}$=-1.35).
\label{simulate_PSD}}
\end{figure}

In the simulations above, we started from the observed 2--10 keV light curves and the best-fit smooth ``softer-and-brighter" trends, and only Poisson errors were added to the expected 2--4 keV and 4--10 keV light curves. Below we perform more general simulations starting from power spectral density  (PSD). 
We use a single powerlaw PSD to simulate exponential light curves (since NGC 4051 is known to show a linear rms-flux relation, e.g., \citealt[][]{Vaughan2011}) 
based on the algorithms of \cite{Uttley2005} and \cite{Emmanoulopoulos2013}.
We use the same PSD shape to simulate two light curves (for 2--4 keV and 4--10 keV respectively).
Nearly identical random numbers are adopted in the simulations to mimic the high correlation between two bands (with a CCF peak value of 0.96).
Two light curves are set to have lengths 1000 times of the observed one shown in Fig. \ref{f1}, then chopped into 1000 segments, each with length and sampling matched to observations.
The simulations were tuned to match the averaged variation amplitude from 1000 segments in each band with the observed one.
The average count rate of each segment is scaled to the observed value so that the effect of added Poisson noise is comparable to that in observed light curves. 
We calculate the SFs and SF ratio from each pair of segments, and derive the corresponding 16\%--84\% percentile ranges of the 1000 segments. 
We simply choose a powerlaw PSD slope of $\alpha$ = -1.6 with which the yielded median SF shape is comparable to the observed ones (upper panel in Fig. \ref{simulate_PSD}).
We find that while the simulated SFs from individual segments show significant scatter (due to red noise leakage and aliasing), the ratio of SFs in two bands shows little deviation from a flat profile
(lower panel in Fig. \ref{simulate_PSD}). This indicates that the SF ratio from two highly correlated light curves,  unlike the SF itself, is barely biased by red noise leakage and aliasing effect.
The observed SF ratio which increases with the timescale clearly contradicts the assumption that two bands have the same PSD shape. 
We further simulate 4--10 keV light curves using PSD slope flatter than 2--4 keV (simply increasing $\alpha$ by 0.25), and find that the observed SF ratio could be qualitatively recovered (Fig. \ref{simulate_PSD}).

\section{Discussion}
Three major scenarios have been proposed to describe/explain the X-rays spectral variabilities in NGC 4051 (and also for other AGNs), including
1) intrinsic spectral variability \citep[the standard model, e.g., spectral pivoting of the powerlaw continuum, e.g.,][]{Uttley2004};
2) variable ionized absorption, in response to X-ray flux variation \citep[e.g.][]{Pounds2004};
3) the combination of a constant reflection component with an intrinsic power-law which is variable in flux but not shape \citep[likely due to light bending effect, e.g.][]{Ponti2006}.
In this work we interpret our discoveries within the standard framework attributing the observed spectral variability entirely to intrinsic variation of the corona, but will also briefly discuss two other scenarios where relevant.  

As we show in Fig. \ref{f1} and Fig. \ref{f2},
we have seen stochastic deviations from the smooth empirical relation
both within and between individual exposures, suggesting variations in the corona geometry or inner structure.  Particularly, we have seen fast deviations at timescales as short as 2 ks. We do not find clear correlation between the deviation and X-ray brightness.
Such quick and stochastic events could not be attributed to the mechanism which is responsible for the general ``softer-when-brighter" trend,  and are possibly due to individual flares within the corona. 
We also present clear evidence that the ``softer-when-brighter" trend is timescale dependent, in the way that the trend is more prominent at longer timescales (from 0.5 ks up to 50 ks). This demonstrates, from a different aspect, that the X-ray spectral shape of NGC 4051 is not solely determined by its brightness.

We remark that in the range of 2 -- 10 keV, the contribution of the reflected continuum and the Fe K fluorescent line is non-negligible, particularly to 4 -- 10 keV. 
A constant reflection component plus a varying powerlaw with constant spectral slope could naturally yield a ``softer-when-brighter" trend, however such an effect alone appears insufficient to explain the observations \citep[e.g.][]{Sobolewska2009,Lamer2003}. We further note that this model,  in which timescales is not involved, would yield a timescale independent  ``softer-when-brighter" trend, inconsistent with observations.
This is also directly confirmed through calculating the structure functions from observations in linear space (simply in unit of count per second), i.e., not affected by constant contamination to the light curves. Considering the reflection component could response and lag behind the intrinsic continuum variation,  we would contrarily expect even stronger ``softer-when-brighter" trends at shorter timescales. 
Attributing the general ``softer-when-brighter" trend in AGNs completely to a varying ionized absorber is also challenging, with strict constraints on the column density, gas density, recombination timescale, location, ionization parameter, and possibly also multi-layer structure of the absorber \citep[e.g.][]{Sobolewska2009,Uttley2004,Pounds2004}. Here we provide a side note 
that the ionized absorber could produce a weaker ``softer-when-brighter" trend at a timescale shorter than the recombination timescale. This qualitatively agrees with our discovery, but quantitative analysis is beyond the scope of this work. 

\subsection{Comparison with previous studies}

We note \cite{Lobban2016,Lobban2018} found in PG 1211+143 and Ark 120 that at short timescales (within individual exposures) the fractional variation amplitude rms  is roughly constant with energy between 2--10 keV, whereas the rms decreases with energy at long timescales (between exposures). 
Such results, derived through comparison between two timescales as a function of energy, are qualitatively consistent with ours (i.e., the ``softer-when-brighter" trend is weaker at a shorter timescale), howbeit our approach provides a straightforward demonstration of the timescale dependency over a broad range of timescale. With our approach, it is also straightforward to identify intervals which deviate from the single empirical trend for future extensive studies. 

There are also studies in literature showing that high frequency X-ray power density spectra (PSD) of AGNs (including NGC 4051) tend to be flatter at higher energy \cite[e.g.][]{Papadakis1995, Vaughan2003a, Vaughan2003b, McHardy2004, Markowitz2005, Vaughan2011, Emmanoulopoulos2016}.
A flatter PSD at higher energy corresponds to a flatter structure function, thus could reproduce a timescale dependency 
of the ``softer-when-brighter" trend (see Fig. \ref{simulate_PSD}).
However, the conclusions in literature were actually derived through comparing the PSD slope in 2--10 keV with those in softer bands but
without correcting the contamination from the soft X-ray excess. 
We stress that the physical origin and variation properties of the soft X-ray excess are known to be different from the powerlaw component \citep[e.g.][]{Jin2017},
and the soft X-ray excess is significant in all the sources in the aforementioned studies. 
\cite{Nandra2001} focused on the powerlaw component, and found the PSD shape
of NGC 7469
is flatter at higher energy. However, their primary discovery is that the PSD shape
in 10--15 keV is flatter than that in 2--4 and 4--10 keV. The PSD shape
in 10-15 keV could be significantly affected by background subtraction (for RXTE
PCA which collects photons with collimator, subtracting background is not as
straightforward as directly imaging X-ray telescopes with mirrors). Actually, between the
two bands with much higher S/N and much lower background contamination (2--4 keV
and 4--10 keV), they did not find statistically significant difference in the PSD
shape.
Meanwhile, \cite{McHardy2004} did split 2--10 keV data of NGC 4051 into two bands (2--5 keV and 5--10 keV) but no statistical significant difference in the PSD slope between two bands was detected (see \citealt{Emmanoulopoulos2016} for a similar study on NGC 7314).
Therefore, there is yet no solid evidence in these studies that the energy dependency of the PSD slope holds for the powerlaw component alone in AGNs.

Flatter PSD at higher energy is also seen in the low/hard and soft/high state in X-ray binaries \citep[e.g.][]{Nowak1999, Grinberg2014}. 
Note the contamination from the thermal X-ray emission from the disk is strong in X-ray binaries in soft/high state.
Indeed, \cite{Zdziarski2005} has shown that the energy dependence of the PSD in the high soft state of GRS 1915+105 could be attributed to
superposition of variability from the less variable disc emission and more variable Comptonization emission.
Meanwhile, while the flatter PSD at higher energies in low/hard states of X-ray binaries is similar to the discovery in this work, the low/hard states of X-ray binaries  are commonly considered as analogs of low luminosity AGNs but not normal AGNs at moderate to high accretion rates \citep[e.g.][]{Markowitz2005},  
thus spectral variability in the low/hard states of X-ray binaries could also have physical origin different from that of the discovery reported in this work.

We conclude that our discoveries of flatter structure function in 4--10 keV (compared with 2--4 keV) in NGC 4051, and subsequently weaker ``softer-when-brighter" trends (of the powerlaw component alone) at shorter timescales, provide new clues to interpret the powerlaw spectral variability and the corona physics in AGNs.

We need to further point out that our investigation of the X-ray spectral variability (the ratio of structure functions) is somehow in analogy to differential photometry, which outperforms absolute photometry in accuracy in time domain studies.
This might be able to explain why \cite{McHardy2004}  did not reveal significant difference in the PSD slope between 2--5 keV and 5--10 keV in NGC 4051, as the light curves and PSDs from two bands were treated as independent to each other.  
In other words, while studying the spectral variations of individual AGNs, a differential approach should be adopted. 
We defer a more extensive comparison between PSD and structure function analyses to a future study, and
focus on the physical interpretation of our results hereinafter.

\subsection{The underlying processes}

Starting from the disc fluctuation model \citep[e.g.][]{Lyubarskii1997}, an inward-propagation scenario has been developed in literature to explain the X-ray spectral-timing variabilities of accreting black holes \citep[e.g.][]{Churazov2001,Kotov2001,Arevalo2006}.  In the model, the inward-propagation of the perturbations from larger radii of an extended corona yields harder X-ray variation lagged behind softer X-ray and flatter PSD in harder band (e.g., see Fig. 9 in \citealt{Kotov2001}). Within this scheme, the ``softer-when-brighter" trend could be attributed to the larger variation amplitude in the softer band, while a flatter PSD in the harder band could reduce the ``softer-when-brighter" trend at shorter timescales as we observed.
However, the inward-propagation model is unable to explain the dynamical/geometrical variations of the corona recently discovered \citep[e.g.][]{Wilkins2015,Alston2020}. Furthermore, the effect of corona temperature has not been considered in the inward-propagation model, which simply assumes softer X-ray emission is produced at larger radii.
Assuming in the scheme the outer corona where softer X-ray emission is produced has lower temperature, the inward-propagation model would yield lower effective corona temperature during the brighter and softer phases, contrary to the ``hotter-when-brighter'' pattern seen in AGNs \citep{Zhang2018}. 
Also note that the studies, which compared the inward-propagation model with AGN observations \citep[e.g.][]{Arevalo2006}, did not either consider the contamination of the soft X-ray excess component. 
Therefore, while the propagation could be at work\footnote{The propagation within the corona might not be inward, but upward as recent MHD simulations suggested \citep{Schnittman2013}.}, it is likely not the dominant mechanism behind the powerlaw spectral variations in AGNs. 

\begin{figure}[ht!]
\plotone{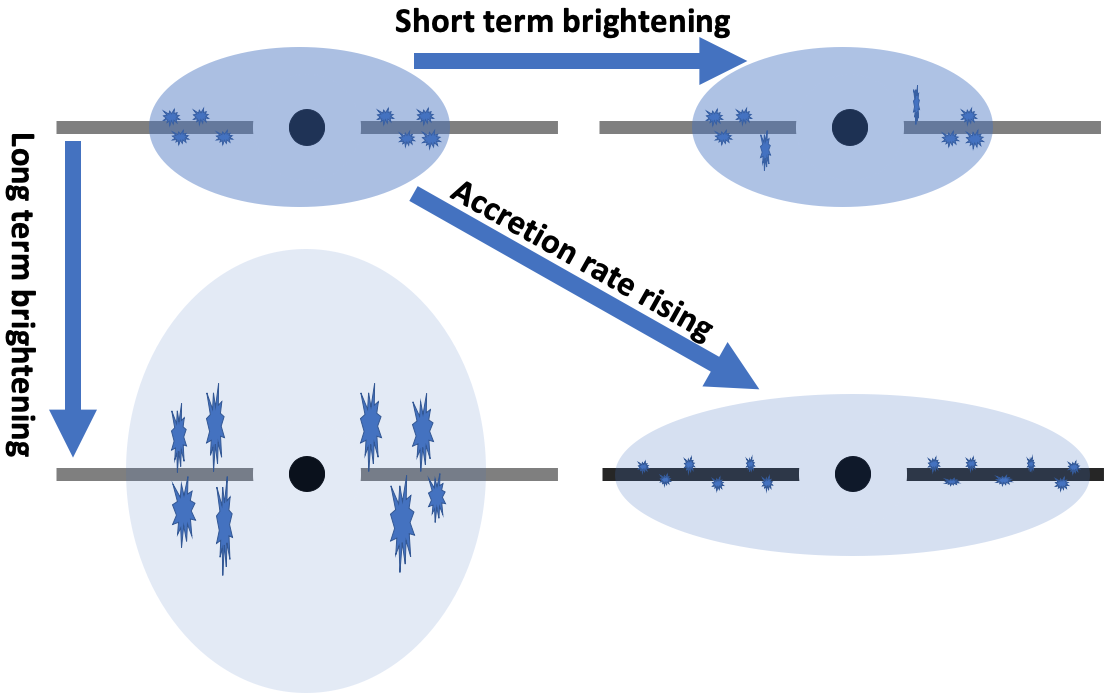}
\caption{A toy scheme (figure not to scale) to describe the X-ray spectral variations in AGNs with moderate to high accretion rates. 
The black hole, the thin accretion disc, the individual flares on top of the disc, and the extended corona are drawn.
Lighter colors of the extended corona signify softer X-ray spectra it produces,
and its size approximately depicts the total power. 
\label{toymodel}}
\end{figure}

Below we show how the findings presented in this work (together with other clues introduced in \S\ref{sec:intro}) could be interpreted  with an alternative scheme as plotted in Fig. \ref{toymodel}, 
in which the propagation is presently neglected. 
In the scheme, the corona has a two-tier structure, including individual flares on top of the accretion disc, embedded in an extended corona
which is heated by the flares,  just like that the solar corona could be heated by solar flare and nano-flares \citep{Parker1988}.
This is in analogy to the solar flares/nano-flares and corona geometry, though the exact underlying physical processes (e.g., corona heating) are yet unclear.
Note other processes which are not yet considered in the current phenomenological scheme,  including those aforementioned (light bending effect, variable ionized absorber) and pair production \citep[e.g.][]{Fabian2015}, may also regulate the spectral variation in AGNs. 

In the scheme rapid variations are dominantly caused by minor and stochastic changes of the individual flares, which merely change the status (geometry, temperature and opacity) of the extended corona, and thus the ``softer-when-brighter" trend is weak. 
On longer timescales, when the flares are significantly stronger, the extended corona is further heated and driven to expand. 
The inflation of the hotter corona leads to smaller opacity (likely also due to higher escape velocity of outflowing electrons),  
thus softer X-ray spectrum could be yielded.
The hotter temperature and/or the higher outflowing velocity are responsible for the higher cutoff energy observed during the brighter and softer phases \citep{Zhang2018}.
Note the correlation between the corona temperature or cutoff energy likely and X-ray flux might be more complicated than a single monotonous function, since more physical processes other than heating and inflation could be involved. For instance, a significantly inflated corona could intercept more seed photons from the disc thus may be contrarily cooled down due to higher cooling efficiency.
Extensive studies of the variations of corona temperature and/or cutoff energy are essential to reveal the physical processes behind the spectral variabilities. 
The slow variations, accompanied with inflation/contraction of the extended corona, could be driven by the variability in the global activity of the inner disc (akin to that of the solar activity, but aperiodic)\footnote{see https://blogs.nasa.gov/sunspot/2018/10/10/solar-cycle-24-in-x-ray-vision/ for an illustration of solar corona inflation/contraction in X-ray images within a solar cycle.}.
Here we refer activity to the strength of the yet unknown processes heating the coronae in AGNs, such as magnetic turbulences, but not the accretion rate.
Note the inflation/contraction describes the geometrical changes of the dominant X-ray emitting region, but not necessarily the outflowing/inflowing velocities of particles in the corona,
thus do not contradict the scenario of outflowing corona \citep[e.g.][]{Liu2014}.
In the scheme, the deviations from the general ``softer-when-brighter" trend could be attributed to the random scattering of the physical properties of individual flares or long term activities. 
The number of the flares should not be too small, to explain the general and rather smooth ``softer-when-brighter" trend, and should not be too large either otherwise the deviations would be smeared out. 

Contrarily,  in case of a higher accretion rate (i.e., on very very long time scales), the inner disc activity is expected to be weaker, as it is known that the X-ray and UV variations anti-correlate with Eddington ratio \citep[e.g.][]{Soldi2014,Kang2018}. Thus a relatively weak extended corona (compared with the total gravitational energy released) could be produced.
The corona cooling would be more efficient since more seed disc photons are produced at higher Eddington ratios and the corona temperature is expected to be lower \citep[see][for marginal observational evidence]{Ricci2018}. The lower temperature (perhaps in combination with lower corona opacity) could then lead to softer X-ray spectra.

The fast deviations/flares could emerge/disappear at timescales as short as $\sim$ 2 ks.  This is still considerably larger than the light travel time ($\sim$ 170 s)
for a 20 $R_g$ ($R_g$ = $GM/c^2$) radius corona in NGC 4051 (with an SMBH mass of 1.7$\pm$0.5 $\times$ 10$^{6}$ M$_{\sun}$, \citealt{Denney2009}). 
As the Compton cooling timescale is expected to be smaller than the light travel time \citep[e.g.][]{Fabian2015}, the observed timescale should reflect
that of the variation of the driven magnetic field.
Next generation X-ray telescopes with much larger photon collection areas would be able to detect even rapider deviations, and fast variations of the physical
parameters (spectral shape, temperature, etc) of the corona.

We finally remark that there are essential differences between our scheme and  a similar two-process model in literature (long term coronal cooling, and short term flares) proposed  by \cite{Nandra2001}. One key difference is that, in our scheme the long term ``softer-when-brighter'' trend is not driven by corona cooling but corona heating and inflation, as recent NuSTAR observations show hints of hotter corona at higher fluxes in individual AGNs \citep{Zhang2018}.  We also show in our scheme that the long term spectral variation in individual AGNs should be intrinsically different from that between AGNs with different accretion rates (see Fig. \ref{toymodel}).

A further note is that the observed rms-flux relation in NCG 4051 \citep[e.g.][]{Vaughan2011}, i.e,  is another important clue to constrain the variability model.
To reproduce the observed linear rms-flux relation, in our model we would expect stronger individual flares during brighter intervals. 
It might be unsurprising if the X-ray brightness in individual AGNs is controlled by the amplitude of inner disc turbulences \citep{Kang2018} but not variation in global accretion rate, 
that stronger turbulences (thus stronger rapid variabilities) are expected during brighter intervals. 
We need to develop our simple toy model into a more concrete one in the future to examine whether it could quantitatively reproduce the observed rms-flux relation (and other X-ray variability behaviors). 
Furthermore,  taking the new aspects presented in this work into consideration, future systematic comparison of the X-ray spectral variability behaviors between AGNs, black hole X-ray binaries and neutron star X-ray binaries  \citep[e.g.][]{Psaltis1999} may yield new clues to the underlying physics.

\section{Summaries}

Using XMM-Newton exposures on NGC 4051, we show that its powerlaw (2--10 keV) spectral variability is more than what an empirical ``softer-when-brighter" trend could describe.
We find intervals which clearly deviate from the empirical relation, both between and within XMM-Newton exposures. The deviations can be seen at timescales as short as $\sim$ 2 ks. 
We further find that the 4 -- 10 keV structure function is flatter than that of 2 -- 4 keV, indicating the ``softer-when-brighter" trend  is gradually weaker at shorter timescales (from 100 ks to 0.5 ks). 
These discoveries directly demonstrate that the powerlaw spectral slope is not solely determined by its brightness. 

We point out that while there are studies in literature finding flatter PSD slopes at higher energies in AGNs,  seemingly consistent with our results (flatter PSD vs flatter structure function), most of such studies did not distinguish the
soft X-ray excess component from the powerlaw emission, and there is yet no clear evidence in literature showing the PSD slopes of the powerlaw component alone in AGNs 
is energy dependent. Our discoveries provide new observational clues to understand the physical nature of the powerlaw spectral variabilities in AGNs.

While the inward-propagation model of an extended X-ray corona \citep[e.g.][]{Kotov2001} could reproduce a flatter PSD in harder band, thus weaker ``softer-when-brighter" trends at shorter timescales,
it is unable to explain other essential observations in literature, including the dynamical/geometrical variations of the corona in AGNs \citep[e.g.][]{Wilkins2015,Alston2020}, and the ``hotter-when-brighter'' pattern recently observed \citep{Zhang2018}. 

In this work we propose a different scheme (Fig. \ref{toymodel}) to interpret the spectral variabilities we observe together with those critical clues reported in literature. 
In the scheme, rapid X-ray variations are attributed to flares/nano-flares on top of the inner accretion disc, which barely alter the spectral slope of the coronal emission.
Meanwhile slow variations, driven by the variability of the global activity (akin to solar activity, not the accretion rate) of the inner region, lead to inflation/contraction of the embedding extended corona.
When more energy is dissipated into the corona,  higher X-ray fluxes are observed, the corona is heated to expand, and the softer 
spectra are produced by the smaller opacity of an inflated corona. 
Contrarily, since a smaller fraction of energy (normalized to the bolometric luminosity) is dissipated into the corona in AGNs with higher Eddington ratios, 
the mechanism behind their softer X-ray spectra should be different (likely due to the weaker activity thus cooler corona in the inner region). 

\section*{Acknowledgement}
We thank the anonymous referees for constructive comments which are helpful to improve the manuscript. 
This work is supported by National Science Foundation of China (grants No. 11421303 $\&$ 11890693)
and CAS Frontier Science Key Research Program (QYZDJ-SSW-SLH006).
We thank Tian Hui and Chen Yao for helpful discussion on solar flares. 

\bibliography{refer}

\begin{thebibliography}{}
\expandafter\ifx\csname natexlab\endcsname\relax\def\natexlab#1{#1}\fi

\bibitem[{{Alston} {et~al.}(2013{\natexlab{a}}){Alston}, {Vaughan}, \&
  {Uttley}}]{Alston2013MNRAS}
{Alston}, W.~N., {Vaughan}, S., \& {Uttley}, P. 2013{\natexlab{a}}, \mnras,
  435, 1511

\bibitem[{{Alston} {et~al.}(2013{\natexlab{b}}){Alston}, {Vaughan}, \&
  {Uttley}}]{Alston2013}
---. 2013{\natexlab{b}}, \mnras, 429, 75

\bibitem[{{Alston} {et~al.}(2020){Alston}, {Fabian}, {Kara}, {Parker},
  {Dovciak}, {Pinto}, {Jiang}, {Middleton}, {Miniutti}, {Walton}, {Wilkins},
  {Buisson}, {Caballero-Garcia}, {Cackett}, {De Marco}, {Gallo}, {Lohfink},
  {Reynolds}, {Uttley}, {Young}, \& {Zogbhi}}]{Alston2020}
{Alston}, W.~N., {Fabian}, A.~C., {Kara}, E., {et~al.} 2020, Nature Astronomy,
  2

\bibitem[{{Ar{\'e}valo} \& {Uttley}(2006)}]{Arevalo2006}
{Ar{\'e}valo}, P., \& {Uttley}, P. 2006, \mnras, 367, 801

\bibitem[{{Cai} {et~al.}(2019){Cai}, {Sun}, {Wang}, {Zhu}, {Gu}, \&
  {Yuan}}]{Cai2019}
{Cai}, Z., {Sun}, Y., {Wang}, J., {et~al.} 2019, Science China Physics,
  Mechanics, and Astronomy, 62, 69511

\bibitem[{{Cai} {et~al.}(2016){Cai}, {Wang}, {Gu}, {Sun}, {Wu}, {Huang}, \&
  {Chen}}]{Cai2016}
{Cai}, Z.-Y., {Wang}, J.-X., {Gu}, W.-M., {et~al.} 2016, \apj, 826, 7

\bibitem[{{Cai} {et~al.}(2018){Cai}, {Wang}, {Zhu}, {Sun}, {Gu}, {Cao}, \&
  {Yuan}}]{Cai2018}
{Cai}, Z.-Y., {Wang}, J.-X., {Zhu}, F.-F., {et~al.} 2018, \apj, 855, 117

\bibitem[{{Churazov} {et~al.}(2001){Churazov}, {Gilfanov}, \&
  {Revnivtsev}}]{Churazov2001}
{Churazov}, E., {Gilfanov}, M., \& {Revnivtsev}, M. 2001, \mnras, 321, 759

\bibitem[{{Connolly} {et~al.}(2016){Connolly}, {McHardy}, {Skipper}, \&
  {Emmanoulopoulos}}]{Connolly2016}
{Connolly}, S.~D., {McHardy}, I.~M., {Skipper}, C.~J., \& {Emmanoulopoulos}, D.
  2016, \mnras, 459, 3963

\bibitem[{{Denney} {et~al.}(2009){Denney}, {Watson}, {Peterson}, {Pogge},
  {Atlee}, {Bentz}, {Bird}, {Brokofsky}, {Comins}, {Dietrich}, {Doroshenko},
  {Eastman}, {Efimov}, {Gaskell}, {Hedrick}, {Klimanov}, {Klimek}, {Kruse},
  {Lamb}, {Leighly}, {Minezaki}, {Nazarov}, {Petersen}, {Peterson},
  {Poindexter}, {Schlesinger}, {Sakata}, {Sergeev}, {Tobin}, {Unterborn},
  {Vestergaard}, {Watkins}, \& {Yoshii}}]{Denney2009}
{Denney}, K.~D., {Watson}, L.~C., {Peterson}, B.~M., {et~al.} 2009, \apj, 702,
  1353

\bibitem[{{di Clemente} {et~al.}(1996){di Clemente}, {Giallongo}, {Natali},
  {Trevese}, \& {Vagnetti}}]{diClemente1996}
{di Clemente}, A., {Giallongo}, E., {Natali}, G., {Trevese}, D., \& {Vagnetti},
  F. 1996, \apj, 463, 466

\bibitem[{{Emmanoulopoulos} {et~al.}(2013){Emmanoulopoulos}, {McHardy}, \&
  {Papadakis}}]{Emmanoulopoulos2013}
{Emmanoulopoulos}, D., {McHardy}, I.~M., \& {Papadakis}, I.~E. 2013, \mnras,
  433, 907

\bibitem[{{Emmanoulopoulos} {et~al.}(2010){Emmanoulopoulos}, {McHardy}, \&
  {Uttley}}]{Emmanoulopoulos2010}
{Emmanoulopoulos}, D., {McHardy}, I.~M., \& {Uttley}, P. 2010, \mnras, 404, 931

\bibitem[{{Emmanoulopoulos} {et~al.}(2016){Emmanoulopoulos}, {McHardy},
  {Vaughan}, \& {Papadakis}}]{Emmanoulopoulos2016}
{Emmanoulopoulos}, D., {McHardy}, I.~M., {Vaughan}, S., \& {Papadakis}, I.~E.
  2016, \mnras, 460, 2413

\bibitem[{{Emmanoulopoulos} {et~al.}(2012){Emmanoulopoulos}, {Papadakis},
  {McHardy}, {Ar{\'e}valo}, {Calvelo}, \& {Uttley}}]{Emmanoulopoulos2012}
{Emmanoulopoulos}, D., {Papadakis}, I.~E., {McHardy}, I.~M., {et~al.} 2012,
  \mnras, 424, 1327

\bibitem[{{Fabian} {et~al.}(2015){Fabian}, {Lohfink}, {Kara}, {Parker},
  {Vasudevan}, \& {Reynolds}}]{Fabian2015}
{Fabian}, A.~C., {Lohfink}, A., {Kara}, E., {et~al.} 2015, \mnras, 451, 4375

\bibitem[{{Galeev} {et~al.}(1979){Galeev}, {Rosner}, \& {Vaiana}}]{Galeev1979}
{Galeev}, A.~A., {Rosner}, R., \& {Vaiana}, G.~S. 1979, \apj, 229, 318

\bibitem[{{Grinberg} {et~al.}(2014){Grinberg}, {Pottschmidt}, {B{\"o}ck},
  {Schmid}, {Nowak}, {Uttley}, {Tomsick}, {Rodriguez}, {Hell}, {Markowitz},
  {Bodaghee}, {Cadolle Bel}, {Rothschild}, \& {Wilms}}]{Grinberg2014}
{Grinberg}, V., {Pottschmidt}, K., {B{\"o}ck}, M., {et~al.} 2014, \aap, 565, A1

\bibitem[{{Grupe} {et~al.}(2010){Grupe}, {Komossa}, {Leighly}, \&
  {Page}}]{Grupe2010}
{Grupe}, D., {Komossa}, S., {Leighly}, K.~M., \& {Page}, K.~L. 2010, \apjs,
  187, 64

\bibitem[{{Haardt} \& {Maraschi}(1991)}]{Haardt1991}
{Haardt}, F., \& {Maraschi}, L. 1991, \apjl, 380, L51

\bibitem[{{Haardt} \& {Maraschi}(1993)}]{Haardt1993}
---. 1993, \apj, 413, 507

\bibitem[{{Iwasawa} {et~al.}(2004){Iwasawa}, {Miniutti}, \&
  {Fabian}}]{Iwasawa2004}
{Iwasawa}, K., {Miniutti}, G., \& {Fabian}, A.~C. 2004, \mnras, 355, 1073

\bibitem[{{Jin} {et~al.}(2017){Jin}, {Done}, \& {Ward}}]{Jin2017}
{Jin}, C., {Done}, C., \& {Ward}, M. 2017, \mnras, 468, 3663

\bibitem[{{Kang} {et~al.}(2018){Kang}, {Wang}, {Cai}, {Guo}, {Zhu}, {Cao},
  {Gu}, \& {Yuan}}]{Kang2018}
{Kang}, W.-y., {Wang}, J.-X., {Cai}, Z.-Y., {et~al.} 2018, \apj, 868, 58

\bibitem[{{Keek} \& {Ballantyne}(2016)}]{Keek2016}
{Keek}, L., \& {Ballantyne}, D.~R. 2016, \mnras, 456, 2722

\bibitem[{{Kotov} {et~al.}(2001){Kotov}, {Churazov}, \& {Gilfanov}}]{Kotov2001}
{Kotov}, O., {Churazov}, E., \& {Gilfanov}, M. 2001, \mnras, 327, 799

\bibitem[{{Lamer} {et~al.}(2003){Lamer}, {McHardy}, {Uttley}, \&
  {Jahoda}}]{Lamer2003}
{Lamer}, G., {McHardy}, I.~M., {Uttley}, P., \& {Jahoda}, K. 2003, \mnras, 338,
  323

\bibitem[{{Liu} {et~al.}(2014){Liu}, {Wang}, {Yang}, {Zhu}, \&
  {Zhou}}]{Liu2014}
{Liu}, T., {Wang}, J.-X., {Yang}, H., {Zhu}, F.-F., \& {Zhou}, Y.-Y. 2014,
  \apj, 783, 106

\bibitem[{{Lobban} {et~al.}(2018){Lobban}, {Porquet}, {Reeves}, {Markowitz},
  {Nardini}, \& {Grosso}}]{Lobban2018}
{Lobban}, A.~P., {Porquet}, D., {Reeves}, J.~N., {et~al.} 2018, \mnras, 474,
  3237

\bibitem[{{Lobban} {et~al.}(2016){Lobban}, {Vaughan}, {Pounds}, \&
  {Reeves}}]{Lobban2016}
{Lobban}, A.~P., {Vaughan}, S., {Pounds}, K., \& {Reeves}, J.~N. 2016, \mnras,
  457, 38

\bibitem[{{Lusso} {et~al.}(2012){Lusso}, {Comastri}, {Simmons}, {Mignoli},
  {Zamorani}, {Vignali}, {Brusa}, {Shankar}, {Lutz}, {Trump}, {Maiolino},
  {Gilli}, {Bolzonella}, {Puccetti}, {Salvato}, {Impey}, {Civano}, {Elvis},
  {Mainieri}, {Silverman}, {Koekemoer}, {Bongiorno}, {Merloni}, {Berta}, {Le
  Floc'h}, {Magnelli}, {Pozzi}, \& {Riguccini}}]{Lusso2012}
{Lusso}, E., {Comastri}, A., {Simmons}, B.~D., {et~al.} 2012, \mnras, 425, 623

\bibitem[{{Lyubarskii}(1997)}]{Lyubarskii1997}
{Lyubarskii}, Y.~E. 1997, \mnras, 292, 679

\bibitem[{{Markowitz} \& {Edelson}(2004)}]{Markowitz2004}
{Markowitz}, A., \& {Edelson}, R. 2004, \apj, 617, 939

\bibitem[{{Markowitz} \& {Uttley}(2005)}]{Markowitz2005}
{Markowitz}, A., \& {Uttley}, P. 2005, \apjl, 625, L39

\bibitem[{{McHardy} {et~al.}(2004){McHardy}, {Papadakis}, {Uttley}, {Page}, \&
  {Mason}}]{McHardy2004}
{McHardy}, I.~M., {Papadakis}, I.~E., {Uttley}, P., {Page}, M.~J., \& {Mason},
  K.~O. 2004, \mnras, 348, 783

\bibitem[{{Miniutti} \& {Fabian}(2004)}]{Miniutti2004}
{Miniutti}, G., \& {Fabian}, A.~C. 2004, \mnras, 349, 1435

\bibitem[{{Nandra} \& {Papadakis}(2001)}]{Nandra2001}
{Nandra}, K., \& {Papadakis}, I.~E. 2001, \apj, 554, 710

\bibitem[{{Nowak} {et~al.}(1999){Nowak}, {Vaughan}, {Wilms}, {Dove}, \&
  {Begelman}}]{Nowak1999}
{Nowak}, M.~A., {Vaughan}, B.~A., {Wilms}, J., {Dove}, J.~B., \& {Begelman},
  M.~C. 1999, \apj, 510, 874

\bibitem[{{Papadakis} \& {McHardy}(1995)}]{Papadakis1995}
{Papadakis}, I.~E., \& {McHardy}, I.~M. 1995, \mnras, 273, 923

\bibitem[{{Parker}(1988)}]{Parker1988}
{Parker}, E.~N. 1988, \apj, 330, 474

\bibitem[{{Peterson}(2001)}]{Peterson2001}
{Peterson}, B.~M. 2001, in Advanced Lectures on the Starburst-AGN, ed.
  I.~{Aretxaga}, D.~{Kunth}, \& R.~{M{\'u}jica}, 3

\bibitem[{{Ponti} {et~al.}(2006){Ponti}, {Miniutti}, {Cappi}, {Maraschi},
  {Fabian}, \& {Iwasawa}}]{Ponti2006}
{Ponti}, G., {Miniutti}, G., {Cappi}, M., {et~al.} 2006, \mnras, 368, 903

\bibitem[{{Pounds} {et~al.}(2004){Pounds}, {Reeves}, {King}, \&
  {Page}}]{Pounds2004}
{Pounds}, K.~A., {Reeves}, J.~N., {King}, A.~R., \& {Page}, K.~L. 2004, \mnras,
  350, 10

\bibitem[{{Psaltis} {et~al.}(1999){Psaltis}, {Belloni}, \& {van der
  Klis}}]{Psaltis1999}
{Psaltis}, D., {Belloni}, T., \& {van der Klis}, M. 1999, \apj, 520, 262

\bibitem[{{Ricci} {et~al.}(2011){Ricci}, {Walter}, {Courvoisier}, \&
  {Paltani}}]{Ricci2011}
{Ricci}, C., {Walter}, R., {Courvoisier}, T.~J.~L., \& {Paltani}, S. 2011,
  \aap, 532, A102

\bibitem[{{Ricci} {et~al.}(2018){Ricci}, {Ho}, {Fabian}, {Trakhtenbrot},
  {Koss}, {Ueda}, {Lohfink}, {Shimizu}, {Bauer}, {Mushotzky}, {Schawinski},
  {Paltani}, {Lamperti}, {Treister}, \& {Oh}}]{Ricci2018}
{Ricci}, C., {Ho}, L.~C., {Fabian}, A.~C., {et~al.} 2018, \mnras, 480, 1819

\bibitem[{{Risaliti} {et~al.}(2009){Risaliti}, {Young}, \&
  {Elvis}}]{Risaliti2009}
{Risaliti}, G., {Young}, M., \& {Elvis}, M. 2009, \apjl, 700, L6

\bibitem[{{Sarma} {et~al.}(2015){Sarma}, {Tripathi}, {Misra}, {Dewangan},
  {Pathak}, \& {Sarma}}]{Sarma2015}
{Sarma}, R., {Tripathi}, S., {Misra}, R., {et~al.} 2015, mnras, 448, 1541

\bibitem[{Schmidt {et~al.}(2012)Schmidt, Rix, Shields, Knecht, Hogg, Maoz, \&
  Bovy}]{Schmidt2012}
Schmidt, K.~B., Rix, H.-W., Shields, J.~C., {et~al.} 2012, The Astrophysical
  Journal, 744, 147

\bibitem[{{Schnittman} {et~al.}(2013){Schnittman}, {Krolik}, \&
  {Noble}}]{Schnittman2013}
{Schnittman}, J.~D., {Krolik}, J.~H., \& {Noble}, S.~C. 2013, \apj, 769, 156

\bibitem[{{Seifina} {et~al.}(2018){Seifina}, {Chekhtman}, \&
  {Titarchuk}}]{Seifina2018}
{Seifina}, E., {Chekhtman}, A., \& {Titarchuk}, L. 2018, \aap, 613, A48

\bibitem[{{Shemmer} {et~al.}(2006){Shemmer}, {Brandt}, {Netzer}, {Maiolino}, \&
  {Kaspi}}]{Shemmer2006}
{Shemmer}, O., {Brandt}, W.~N., {Netzer}, H., {Maiolino}, R., \& {Kaspi}, S.
  2006, \apjl, 646, L29

\bibitem[{{Sobolewska} \& {Papadakis}(2009)}]{Sobolewska2009}
{Sobolewska}, M.~A., \& {Papadakis}, I.~E. 2009, \mnras, 399, 1597

\bibitem[{{Soldi} {et~al.}(2014){Soldi}, {Beckmann}, {Baumgartner}, {Ponti},
  {Shrader}, {Lubi{\'n}ski}, {Krimm}, {Mattana}, \& {Tueller}}]{Soldi2014}
{Soldi}, S., {Beckmann}, V., {Baumgartner}, W.~H., {et~al.} 2014, \aap, 563,
  A57

\bibitem[{{Sun} {et~al.}(2014){Sun}, {Wang}, {Chen}, \& {Zheng}}]{Sun2014}
{Sun}, Y.-H., {Wang}, J.-X., {Chen}, X.-Y., \& {Zheng}, Z.-Y. 2014, \apj, 792,
  54

\bibitem[{{Taylor} {et~al.}(2003){Taylor}, {Uttley}, \& {McHardy}}]{Taylor2003}
{Taylor}, R.~D., {Uttley}, P., \& {McHardy}, I.~M. 2003, \mnras, 342, L31

\bibitem[{{Tortosa} {et~al.}(2018){Tortosa}, {Bianchi}, {Marinucci}, {Matt}, \&
  {Petrucci}}]{Tortosa2018}
{Tortosa}, A., {Bianchi}, S., {Marinucci}, A., {Matt}, G., \& {Petrucci}, P.~O.
  2018, \aap, 614, A37

\bibitem[{{Uttley}(2006)}]{Uttley2006}
{Uttley}, P. 2006, Astronomical Society of the Pacific Conference Series, Vol.
  360, {The Relationship Between Optical and X-ray Variability in Seyfert
  Galaxies}, ed. C.~M. {Gaskell}, I.~M. {McHardy}, B.~M. {Peterson}, \& S.~G.
  {Sergeev}, 101

\bibitem[{{Uttley} {et~al.}(2005){Uttley}, {McHardy}, \&
  {Vaughan}}]{Uttley2005}
{Uttley}, P., {McHardy}, I.~M., \& {Vaughan}, S. 2005, \mnras, 359, 345

\bibitem[{{Uttley} {et~al.}(2004){Uttley}, {Taylor}, {McHardy}, {Page},
  {Mason}, {Lamer}, \& {Fruscione}}]{Uttley2004}
{Uttley}, P., {Taylor}, R.~D., {McHardy}, I.~M., {et~al.} 2004, \mnras, 347,
  1345

\bibitem[{{Vasudevan} {et~al.}(2009){Vasudevan}, {Mushotzky}, {Winter}, \&
  {Fabian}}]{Vasudevan2009}
{Vasudevan}, R.~V., {Mushotzky}, R.~F., {Winter}, L.~M., \& {Fabian}, A.~C.
  2009, \mnras, 399, 1553

\bibitem[{{Vaughan} \& {Fabian}(2003)}]{Vaughan2003a}
{Vaughan}, S., \& {Fabian}, A.~C. 2003, \mnras, 341, 496

\bibitem[{{Vaughan} {et~al.}(2003){Vaughan}, {Fabian}, \&
  {Nandra}}]{Vaughan2003b}
{Vaughan}, S., {Fabian}, A.~C., \& {Nandra}, K. 2003, \mnras, 339, 1237

\bibitem[{{Vaughan} {et~al.}(2011){Vaughan}, {Uttley}, {Pounds}, {Nandra}, \&
  {Strohmayer}}]{Vaughan2011}
{Vaughan}, S., {Uttley}, P., {Pounds}, K.~A., {Nandra}, K., \& {Strohmayer},
  T.~E. 2011, \mnras, 413, 2489

\bibitem[{{Wang} {et~al.}(2004){Wang}, {Watarai}, \& {Mineshige}}]{Wang2004}
{Wang}, J.-M., {Watarai}, K.-Y., \& {Mineshige}, S. 2004, \apjl, 607, L107

\bibitem[{{Wilkins} {et~al.}(2015){Wilkins}, {Gallo}, {Grupe}, {Bonson},
  {Komossa}, \& {Fabian}}]{Wilkins2015}
{Wilkins}, D.~R., {Gallo}, L.~C., {Grupe}, D., {et~al.} 2015, \mnras, 454, 4440

\bibitem[{{Yang} {et~al.}(2015){Yang}, {Xie}, {Yuan}, {Zdziarski},
  {Gierli{\'n}ski}, {Ho}, \& {Yu}}]{Yang2015}
{Yang}, Q.-X., {Xie}, F.-G., {Yuan}, F., {et~al.} 2015, \mnras, 447, 1692

\bibitem[{{Zdziarski} {et~al.}(2005){Zdziarski}, {Gierli{\'n}ski}, {Rao},
  {Vadawale}, \& {Miko{\l}ajewska}}]{Zdziarski2005}
{Zdziarski}, A.~A., {Gierli{\'n}ski}, M., {Rao}, A.~R., {Vadawale}, S.~V., \&
  {Miko{\l}ajewska}, J. 2005, \mnras, 360, 825

\bibitem[{{Zdziarski} {et~al.}(1995){Zdziarski}, {Johnson}, {Done}, {Smith}, \&
  {McNaron-Brown}}]{Zdziarski1995}
{Zdziarski}, A.~A., {Johnson}, W.~N., {Done}, C., {Smith}, D., \&
  {McNaron-Brown}, K. 1995, \apjl, 438, L63

\bibitem[{{Zhang} {et~al.}(2018){Zhang}, {Wang}, \& {Zhu}}]{Zhang2018}
{Zhang}, J.-X., {Wang}, J.-X., \& {Zhu}, F.-F. 2018, \apj, 863, 71

\bibitem[{Zhu {et~al.}(2016)Zhu, Wang, Cai, \& Sun}]{Zhu2016}
Zhu, F.-F., Wang, J.-X., Cai, Z.-Y., \& Sun, Y.-H. 2016, The Astrophysical
  Journal, 832, 75

\bibitem[{{Zhu} {et~al.}(2018){Zhu}, {Wang}, {Cai}, {Sun}, {Sun}, \&
  {Zhang}}]{Zhu2018}
{Zhu}, F.-F., {Wang}, J.-X., {Cai}, Z.-Y., {et~al.} 2018, \apj, 860, 29

\end{thebibliography}

\end{document}